\documentclass[aps,prd,preprint,groupedaddress,amsmath,amssymb,floats,nofootinbib,tightenlines,superscriptaddress]{revtex4}
\usepackage{graphicx}
\usepackage{units}
\usepackage[pdfborder={0 0 0}]{hyperref}
\usepackage{rotating}

\DeclareMathOperator{\tr}{tr}

\begin{document}

\newcommand{\figref}[1]{Fig.~\ref{#1}}
\newcommand{\tabref}[1]{Tab.~\ref{#1}}
\newcommand{\secref}[1]{Section \ref{#1}}
\newcommand{\appref}[1]{Appendix \ref{#1}}


\title{Exploring compactified HEIDI models at the LHC}

\begin{flushright}
\end{flushright}

\author{Neil D. Christensen}
\email[]{neilc@pitt.edu}
\affiliation{
PITTsburgh Particle physics, Astrophysics and Cosmology Center, \\ 
Department of Physics and Astronomy, University of Pittsburgh, \\
Pittsburgh, PA 15260, USA
\vspace{1ex}}
\author{Benjamin Fuks}
\email[]{benjamin.fuks@iphc.cnrs.fr}
\affiliation{%
Institut Pluridisciplinaire Hubert Curien/D\'epartement Recherche Subatomique\\
Universit\'e de Strasbourg/CNRS-IN2P3 \\
23 Rue du Loess, F-67037 Strasbourg, France
\vspace{1ex}}
\author{J\"urgen Reuter}
\email[]{juergen.reuter@desy.de}
\affiliation{%
DESY, Theory Group, Bldg. 2a, Notkestr. 85,
D-22603 Hamburg, Germany
\vspace{1ex}}
\author{Christian Speckner}
\email[]{christian.speckner@physik.uni-freiburg.de}
\affiliation{%
Albert-Ludwigs-Universit\"at Freiburg\mbox{,} Physikalisches Institut\\
Hermann-Herder-Stra\ss{}e 3, 79104 Freiburg, Germany
\vspace{1ex}}
\begin{abstract}
  Models with multi-scalar Higgs sectors inspired by a 
  higher-dimensional setup are interesting alternatives to the
  Standard Model because, although they have a Higgs sector which
  gives mass to the W and Z gauge bosons as well as the SM fermions,
  this Higgs sector is potentially undiscoverable at the Large
  Hadron Collider or shows considerable deviations from the Standard
  Model Higgs sector. We investigate a compactified version of such
  models and study its phenomenology in the ``golden'' four-lepton
  channel at the LHC in areas of parameter space compatible with
  electroweak precision observables . 
\end{abstract}

\maketitle

\section{Introduction}
The Standard Model (SM) of particle physics provides a successful description of all
experimental high-energy data to date. 
However, despite its success, many fundamental questions remain unanswered, 
such as the origin of electroweak symmetry breaking, the nature of
neutrino masses, the large hierarchy between the electroweak and the
Planck scales and the origins of dark matter and the cosmological constant. 
Attempts to address these questions have led to a wide range of new physics
theories, most of them predicting new phenomena at the TeV scale. 
The Large Hadron Collider (LHC) at CERN is currently probing this
scale and will hopefully make it possible to discover, constrain
and/or exclude some of the proposed theories beyond the Standard Model (BSM).

The Higgs boson as the last piece of the SM has been the searched for
over several decades now. After the unsuccessful search for it at the
LEP $e^+e^-$ collider as well as the Tevatron (though there are some
hints in a slight excess now in a broad mass range from 120-140 Gev)
several models that explain electroweak symmetry breaking (EWSB) as
well as unitarization of tree 
level scattering amplitudes have been constructed that either work
without a scalar state being present or sport invisible or
undetectable decays of those states. One type of these models are the so-called
HEIDI models~\cite{vanderBij:2006ne,vanderBij:2007um,vanderBij:2011wy}
(an onomatopoeic version of the  
acronym for higher dimensions), based on an older model presented
in~\cite{Hill:1987ea}. The basic idea of this class of models is that
the Higgs field mixes with a certain number of inert (singlet) fields,
the number of which can be increased until they become a continuum. This admixture
with such a set or continuum of fields can distort the properties of
the physical Higgs particle(s), i.e. its production cross sections,
its decay width and branching ratios. At the moment, there seems to be
some evidence for a SM-like Higgs boson around 125 GeV, which could
have modified couplings to SM particles, as the diphoton rate seems to
be rather enhanced, while the $WW^*$ decay seems to be suppressed with
respect to the SM. Recently, the aforementioned models have been
modified in order to precisely explain these deviations from the SM
values~\cite{vanderBij:2012ci}. 

An interesting feature of this type of model is that it may hide the source of
electroweak symmetry breaking from discovery at the LHC.  We know that, if 
the electroweak symmetry is broken explicitly in the Lagrangian (rather than
 through an asymmetric vacuum), the scattering of longitudinal electroweak 
vector bosons grows quadratically with energy and exceeds the
perturbative unitarity bound at just over a TeV.  As a result, we know
that some new physics must appear below that scale in order to
unitarize the theory.  The common assumption is that 
we cannot fail to discover this new physics at the LHC.  HEIDI type models,
however, have areas of parameter space which are consistent with
precision constraints, but may have such a broad signal as to be
undetectable at the LHC with current methods
\cite{vanderBij:2006ne,vanderBij:2007um}. 

In this work we will be examining a HEIDI type model which includes an
infinite but discrete set of additional scalar resonances. The model is realized
by means of a spatial and compact extra dimension with the SM being confined to a
four-dimensional brane within the five-dimensional spacetime. The only field which
is allowed to propagate into the bulk of the extra dimension is a singlet scalar
field coupled to the scalar sector of the SM through a trilinear coupling. After
Kaluza-Klein (KK) reduction, we obtain an infinite number of new scalar fields
mixing with the SM Higgs boson.  Depending on the point in the parameter space,
many KK states could be well below the TeV scale, and the emerging collider
phenomenology of the scalar sector can thus be very different from the SM. In
order to assess the impact of the modified scalar sector on the Higgs searches,
we study the so-called golden channel gluon fusion process.

The outline of this paper is as follows: in
Sec.~\ref{sec:heidi_review} we present our class of HEIDI models
including their Lagrangian and derive the mass eigenvalues and
eigenstates of the scalar sector together with its couplings to the SM
particles. We discuss the parameter space of HEIDI models and examine
different characteristic incarnations of the scalar sector depending
on the hierarchies in the parameter space. The simulation of such
models is potentiall very demanding as the number of physical particles in the
spectrum can be very large and depends on the point in parameter space, and our way
of tackling these technical difficulties is discusse in
Sec.~\ref{sec:implementation}. In Sec.~\ref{sec:golden}, our results
for the golden channel of a Higgs boson decaying into $ZZ^*$ into four
leptons at the LHC are shown, which is (if the Higgs mass is not too
close to the LEP limit) the cleanest and most background-free
channel. Finally, we conclude.

\section{The model}
\label{sec:heidi_review}

\subsection{Lagrangian and spectrum}
\label{sec:heidi-spectrum}

The compact HEIDI model discussed in this paper is a representative of the
class of models suggested in Ref.~\cite{vanderBij:2006ne}. The model is a
renormalizable
extension of the Standard Model formulated on Minkowski spacetime plus an additional
spatial extra dimension compactified on a circle with radius $R$ and denoted in the
 following by the coordinate $y$.
In this model the Standard Model is confined to the $y=0$ brane,
with the Higgs sector supplemented by an additional scalar gauge-singlet field
$\Omega(x,y)$ which
is the only field allowed to propagate in the bulk of the extra dimension. The
action of the scalar sector is chosen\footnote{We omit any self-interactions
of $\Omega$ in the same spirit as Ref.~\cite{vanderBij:2006ne} as they would
spoil renormalizability.} to be
\begin{multline}
S = \underbrace{
   \int d^4 x\;
   \left(\frac{1}{2}\left(D_\mu\Phi\right)^\dagger D^\mu\Phi +
   \frac{\mu^2}{2}\Phi^\dagger \Phi -
   \frac{\lambda}{4}\left(\Phi^\dagger\Phi\right)^2 \right)
}_{S_\Phi}  \\ + \; \; \underbrace{
   \int d^4 x\;\int_0^{2\pi R} dy\;
   \left(\frac{1}{2}\partial_a \Omega\partial^a \Omega -
   \frac{m_b^2}{2}\Omega^2 + g\Phi^\dagger\Phi \Omega\delta(y)\right)
}_{S_\Omega} \ ,
\label{equ:heidi-s5d}
\end{multline}
with $\Phi$ being the Standard Model Higgs doublet (for later
convenience we use a non-canonical normalization of $\Phi$) and
$S_\Phi$ the action describing the SM Higgs sector. From  Eq.~\eqref{equ:heidi-s5d},
it is evident that the only difference with respect to the SM is the additional
sector $S_\Omega$, which contains a trilinear coupling $g$ between the Higgs field
and the new scalar singlet $\Omega$.

In order to work out the equivalent four-dimensional theory, we decompose the new
scalar field in terms of its KK excitations\footnote{The sine modes in the KK
decomposition decouple from the SM,  so we omit them.}
\begin{equation}
\Omega(x,y) = \sum_{k = 0}^{\infty}\frac{1}{N_k\sqrt{R}} \Omega_k(x)\cos\frac{k}{R}y\,,
\qquad\text{with}\qquad
N_k = \sqrt{\left(1+\delta_{k0}\right)\pi} \ .
\label{equ:heidi-kkdec}
\end{equation}
Inserting  Eq.~\eqref{equ:heidi-kkdec} into  Eq.~\eqref{equ:heidi-s5d}, we obtain
the four-dimensional action
\begin{equation}
S_\Omega = \int d^4x\;\sum_{k=0}^{\infty}\left(
   \frac{1}{2}\partial_\mu \Omega_k\partial^\mu \Omega_k -
   \frac{1}{2}m_k^2\Omega_k^2 + \frac{g}{N_k\sqrt{R}}\Phi^\dagger\Phi \Omega_k
\right)\ ,
\label{equ:heidi-s4d}
\end{equation}
where the masses of the KK modes are given by%
\begin{equation*}
m_k^2 = m_b^2 + \frac{k^2}{R^2} \ .
\end{equation*}

Like in the SM, spontaneous symmetry breaking occurs if $\mu^2>0$ (see
\appref{app:heidi-vev} for a detailed discussion of the potential and the vacuum
expectation values). Rotating to
unitarity gauge and shifting the scalar fields by their vacuum expectation
values (vev), we get
\begin{equation}
\Phi(x)_\text{unit.} = \begin{pmatrix}0 \\ v + h(x)\end{pmatrix}
\qquad,\qquad \Omega_k(x) = w_k + \omega_k(x) \ ,
\label{equ:heidi-shifts}
\end{equation}
where the vevs are given by
\begin{equation}
v = \frac{\mu}{\sqrt{\lambda - \alpha}} \;\left(\text{with}\;\;
\alpha = \frac{g^2}{m_b}\coth\left(m_b R\pi\right)\right)
\qquad,\qquad w_k = \frac{gv^2}{\sqrt{R}N_k m_k^2} \ . 
\label{equ:heidi-vevs}
\end{equation}
Defining the Higgs mass in the usual way,
\begin{equation}
m_h = \sqrt{2\lambda} \, v \ , 
\label{equ:heidi-mh}
\end{equation}
the four-dimensional Lagrangian for the scalar fields, expanded around the
correct vacuum, reads
\begin{multline}
\mathcal{L}_\text{scalar} = \frac{1}{2} \partial^\mu h\partial_\mu h -
   \frac{1}{2} m_h^2h^2 -
   \lambda v h^3 - \frac{\lambda}{4} h^4 \\ + \sum_{k=0}^{\infty} \left(
   \frac{1}{2}\partial_\mu\omega_k\partial^\mu\omega_k -
   \frac{1}{2}m_k^2\omega_k^2 + \frac{2gv}{N_k \sqrt{R}}h\omega_k +
   \frac{g}{N_k \sqrt{R}}h^2 \omega_k\right) 
\ .
\label{equ:heidi-lphys}
\end{multline}
In the equation above, we have omitted constant terms as well as all the terms
involving gauge bosons included in the covariant derivative of
Eq.~\eqref{equ:heidi-s5d} (which are unchanged with respect to the SM).
Note that  Eq.~\eqref{equ:heidi-vevs} implies a condition on the quartic Higgs
coupling $\lambda$, the trilinear coupling $g^2$, the bulk mass $m_b$
and the compactification scale $R^{-1}$,
\begin{equation}
\lambda > \alpha = \frac{g^2}{m_b}\coth\left(m_b R\pi\right) \ .
\label{equ:heidi-stability}
\end{equation}
As shown in \appref{app:heidi-vev}, violation of Eq.~\eqref{equ:heidi-stability}
leads to an unstable theory with a potential which is not bounded from
below\footnote{With the exception $\lambda=\alpha$ and $\mu^2<0$, in which
case the potential is bounded and has a single minimum at $v=w_k=0$.}.

Owing to the trilinear coupling between $\Phi$ and $\Omega$ in the original
action, Eq.~\eqref{equ:heidi-lphys} contains a mixing between $h$ and the
$\omega_k$ which therefore do not correspond to mass eigenstates.
While this mixing is resummed to  obtain an effective Higgs
propagator in Ref.\ \cite{vanderBij:2006ne}, we here take a different approach and
diagonalize the mass matrix explicitly. This calculation leads to the mass eigenvalues
$\lambda_k$ as the zeroes of a transcendental function 
(see \appref{app:heidi-wvs} for details),
\begin{equation}
f(\lambda_k^2) = m_h^2 + \frac{2g^2v^2}{\sqrt{\lambda_k^2 - m_b^2}}\cot\left(
   R\pi\sqrt{\lambda_k^2 - m_b^2}\right) - \lambda_k^2 
 = 0 \ .
\label{equ:heidi-mtrans}
\end{equation}
It is easy to see from monotony arguments that the spectrum consists of
exactly one mode $\lambda_k^2$ in each interval
\begin{equation}\label{equ:heidi-scalar-spectrum}
m_b^2 + \frac{(k - 1)^2}{R^2} \;<\; \lambda_k^2 \;<\; m_b^2 + \frac{k^2}{R^2}
\qquad\text{with}\qquad k \ge 1 \ , 
\end{equation}
as well as an additional mode $\lambda_0^2$ below $m_b^2$.

\subsection{SM couplings and parameter space}
\label{sec:heidi-parspace}

The mass eigenstates $\phi_k$ belonging to the mass eigenvalues $\lambda_k^2$
are linear combinations of the fields $h$ and $\omega_k$,
\begin{equation}
\phi_k = \xi_0^k h + \sum_{l=1}^\infty \xi_l^k \omega_{l-1} \ .
\label{equ:heidi-xidef}
\end{equation}
Using the hermiticity of the mass matrix, we can invert this relation to
derive the relations between $h$ and the $\omega_k$ in terms of the mass
eigenstates
\begin{equation}
h = \sum_{l=0}^\infty \xi^l_0 \phi_l \qquad,\qquad
\omega_k = \sum_{l=0}^\infty \xi^l_{k+1} \phi_l
\label{equ:heidi-fieldexp} \ . 
\end{equation}
From  Eq.~\eqref{equ:heidi-fieldexp}, it is clear that the SM fields couple to the
physical fields just like they would couple to the SM Higgs boson, but with a factor
$\xi^l_0$ for each $\phi_l$ leg at the vertex. As the couplings among the
scalar fields are slightly more complicated, we have moved the detailed Feynman 
rules to \appref{app:heidi-fr}.

According to Eq.~\eqref{equ:heidi-s5d}, Eq.~\eqref{equ:heidi-vevs} and
Eq.~\eqref{equ:heidi-mh}, the model can be parameterized by the SM parameters $m_h$
and $v$ together with the bulk mass $m_b$, the trilinear coupling $g^2$ and the
compactification scale $R^{-1}$, which is the parameterization that we will
adopt for
the rest of this work. However, let us note that, while $v$ at tree level has
the usual relation to the mass of the $W$ boson $m_W$,
\begin{equation*}
v = 2\frac{m_W\sin\theta_W}{e} \ ,
\end{equation*}
with $\theta_W$ being the weak mixing angle and $e$ the electromagnetic
coupling,
$m_h$ generally does not coincide with any of the masses of the
physical scalar modes.

\begin{figure}
\centerline{\begin{tabular}{ccc}
\includegraphics[angle=270,width=0.48\textwidth]{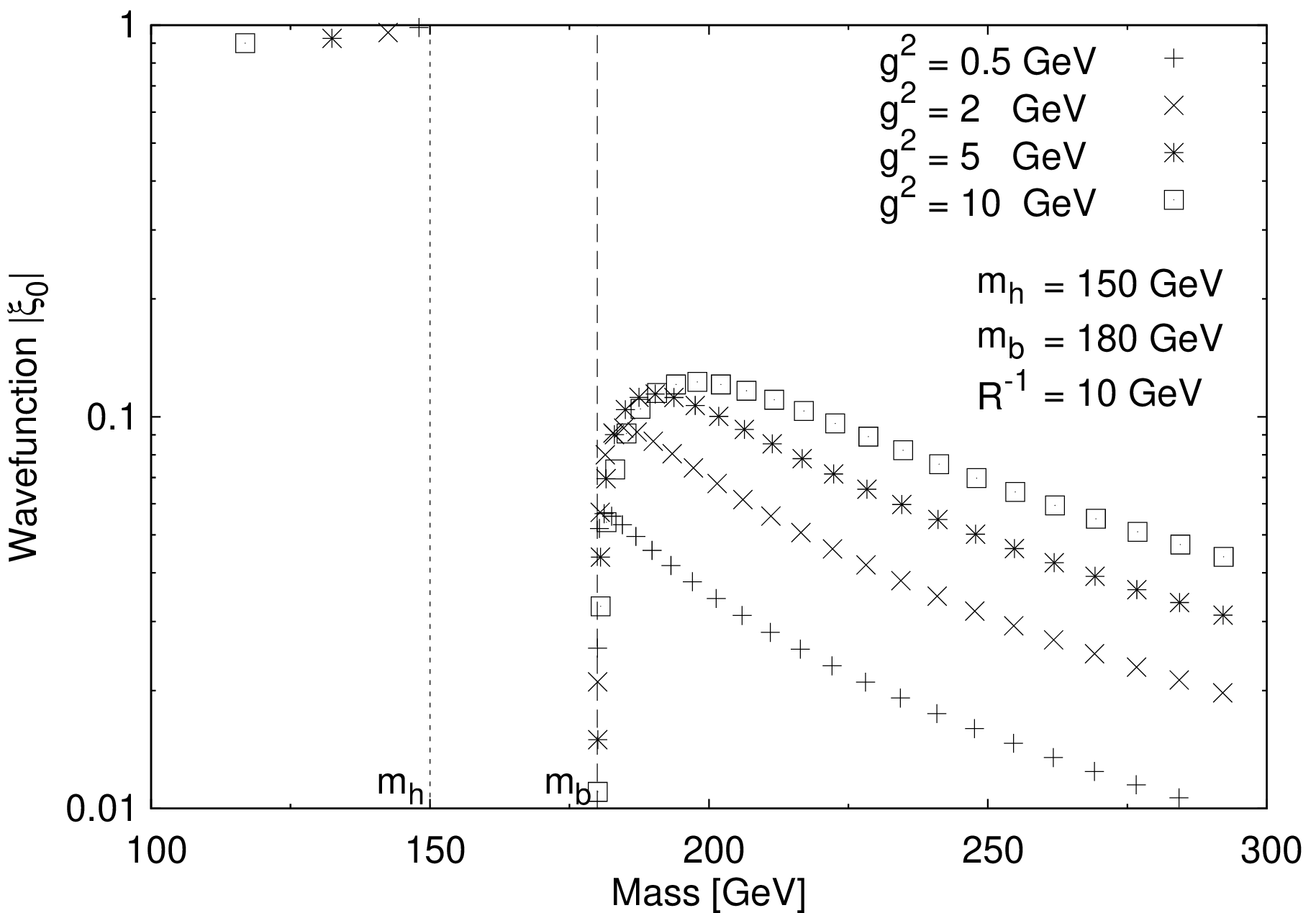} &
\hspace{2ex} &
\includegraphics[angle=270,width=0.48\textwidth]{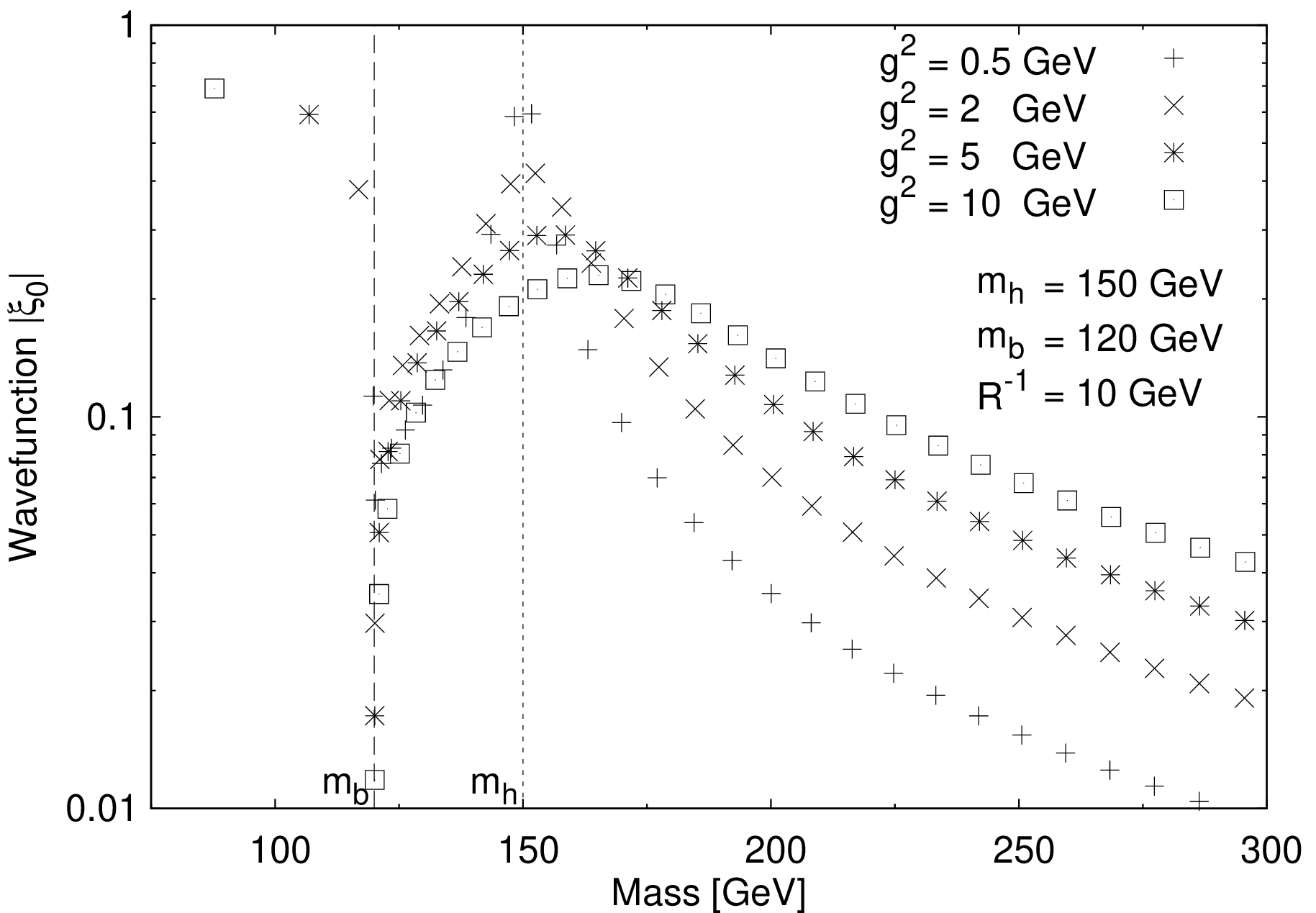}
\end{tabular}}
\caption{Compact HEIDI spectra for the two different scenarios referred to in
the text below. The respective values of $m_h$ and $m_b$ are marked with
vertical lines.}
\label{fig:heidi-spectrum}
\end{figure}
On closer examination of the scalar spectrum, two very different cases can be
distinguished and are shown in \figref{fig:heidi-spectrum} where
$\left|\xi^k_0\right|$ (which controls the couplings to the SM particles) is
displayed as a function of the
mass of the modes\footnote{These figures are closely related to the K\"all\'en-Lehmann
spectral density of $h$ which is given by
$\rho(s) = \sum_{k=0}^\infty \left(\xi^k_0\right)^2\delta\left(s-\lambda_k^2\right)$.}.
If $m_h$ lies below $m_b$ (left
hand side of \figref{fig:heidi-spectrum}), we find that the lightest
mode, which always lies below $m_b$ as argued above, has
$\left|\xi^0_0\right|$ close to one. The other modes form a tower starting at
$m_b$ with the couplings of this
``would-be continuum'' being smaller by about an order
of magnitude and dropping off for higher masses. Increasing $g^2$ pushes
the mass of the lowest mode towards zero and enhances the couplings of
the would-be continuum, while the mass of the lowest mode approaches
$m_h$ in the limit $g^2\rightarrow 0$ with
$\left|\xi^0_0\right|\rightarrow 1$, and the other modes decouple from the SM.

If $m_h$ lies above $m_b$ (right hand side of
\figref{fig:heidi-spectrum}), the situation looks similar for large
values of $g^2$. However, if we decrease this parameter, the lightest
mode moves towards $m_b$ with its coupling to the SM decreasing. At the
same time, a peak in the would-be continuum is appearing at $m_h$.
In the limit $g^2\rightarrow 0$, this peak evolves into a single mode
with a SM coupling of one, while the other modes (this time including
the lowest one) decouple. 

For both cases, the asymptotic spacing between the modes is given by the
compactification scale $R^{-1}$. In the limit of large $R$,
the modes in the would-be continuum approach each other, while their
couplings to the SM go to zero.


\subsection{Considerations regarding simulations}

A thorough study of the phenomenology of compact HEIDI requires its 
implementation into an event generator and the simulation of the different
Higgs production channels within the HEIDI scenario. Considering that all relevant
processes feature at least a four-particle final state already at the parton level
and taking into account that the number of Feynman diagrams is multiplied by
the number of propagating scalar fields, speed is a critical issue here. Therefore,
as the parton-level event generator {\sc Whizard}~\cite{Kilian:2007gr,Moretti:2001zz} 
can handle final states with six to eight particles in the SM and in the Minimal
Supersymmetric Standard Model
(MSSM)~\cite{Hagiwara:2005wg,Kalinowski:2008fk,Robens:2008sa}, and has
been used especially for 
many alternative models of
EWSB~\cite{Ohl:2008ri,Kilian:2006eh,Beyer:2006hx,Kilian:2004pp}, it is a good 
choice for the investigation 
of compact HEIDI models as well.

For the actual implementation of the model, the tower of scalar fields must
be truncated at a cutoff scale $\Lambda$.
However, if we choose the bulk mass $m_b$ to be of the order of the weak 
scale and $R^{-1}$ is small,  then it is easy to see from
Eq.~\eqref{equ:heidi-scalar-spectrum} that the spectrum will contain a
large number of scalar fields below $\Lambda$.
As a consequence, even though at the 
Lagrangian level the HEIDI model is a straightforward deformation of the 
scalar sector of the SM, consisting of the simple addition of a single five-dimensional
scalar singlet, the implementation of the model into {\sc Whizard}
(as well as into any other matrix element generator) will result in the addition of 
a very large number of particles and interactions to the existing SM 
implementation of the code. In addition, the number of fields below the cutoff
depends on the parameters of the model. We therefore choose to follow the approach 
introduced in Ref.~\cite{Christensen:2009jx} and implement the model using 
the newly developed {\sc FeynRules} interface to {\sc Whizard}~\cite{Christensen:2010wz}.
In the next section we briefly outline some of the details of the
implementation using that interface.

\section{Prerequesites for Simulation}
\label{sec:implementation}

In this section we describe the implementation of the compact HEIDI model
of \secref{sec:heidi_review} into {\sc FeynRules}. We first explain
how the numeric spectrum and widths of the scalar fields are obtained for a
given point in parameter space, while the actual {\sc FeynRules} implementation
is presented afterwards.  The {\sc FeynRules} model files are available for
download from the {\sc FeynRules} model database located at
\begin{quotation}
\texttt{https://server06.fynu.ucl.ac.be/projects/feynrules/wiki/ModelFiles}
\end{quotation}

\subsection{Spectrum and widths}

According to Eq.~\eqref{equ:heidi-mtrans}, the masses $\lambda_k$ of the physical
scalar modes are encoded as the zeroes of a transcendental function
$f(\lambda_k^2)$. As no analytic expression for these roots exists, a procedure
for their numerical determination is necessary.

As has already been observed in \secref{sec:heidi-spectrum},
the lowest mode is always located in the interval between $0$ and $m_b$, and the
would-be continuum which starts at $m_b$ has exactly one mode in each of the
intervals of Eq.~\eqref{equ:heidi-scalar-spectrum}.
As $f(\lambda^2)$ is monotonous between the
boundaries of the intervals, it is possible to calculate
the masses by the bisection method: for each mode, we start in the middle of the
respective interval and then, depending on the sign of $f$, split the interval
we know to contain a root in two until sufficient precision has been reached. This
is guaranteed to find all modes, starting from the lowest one.

Once the mass of a mode $\lambda_k$ has been determined, the mixing parameters
$\xi^k_i$ defined in Eq.~\eqref{equ:heidi-xidef} can be calculated as functions of
$\lambda_k$ through application of Eq.~\eqref{equ:heidi-evect1} and
Eq.~\eqref{equ:heidi-evect2}. From the mixing parameters, the trilinear and quartic
scalar self-couplings can be derived according to the Feynman rules given in
\appref{app:heidi-fr}.
We have implemented the calculation of the masses, mixings and self-couplings
into a spectrum calculator. While this is included in the model file
for {\sc FeynRules}, it is also available as a stand-alone
{\sc Mathematica} package.

In addition to masses and couplings, the widths of the scalar modes are
required for simulating collider cross sections. As the couplings of the
$\phi_k$ to the SM particles are identical to the respective Higgs couplings up
to factors of $\xi^k_0$ (\emph{cf.} \appref{app:heidi-fr}), their decay widths into SM
particles can be approximated\footnote{There is a small correction
  coming from diagrams involving both SM particles and other scalar fields
  $\phi_k$.} by the corresponding Higgs widths, scaled by
factors of $\left|\xi^k_0\right|^2$.
We have chosen to take the necessary SM Higgs width from {\sc Hdecay}
\cite{Djouadi:1997yw}. In order to avoid a run-time dependence on the {\sc
Hdecay} package,  we have parametrized the Higgs mass dependence of its output
in a dedicated {\sc Mathematica} module (which is also contained within the
{\sc FeynRules} model file).  As the presence
of thresholds spoils any simple ansatz for fitting, we parameterize the width
by a grid of 100 points for Higgs masses below $\unit[500]{GeV}$ (with linear
interpolation between the grid points), while an
interpolating polynomial of degree 6 is used above $\unit[500]{GeV}$. Comparison
shows that the result approximates the corresponding {\sc Hdecay} output well
at least up to $\unit[4]{TeV}$.

Using this parameterization, our
spectrum calculator is able to automatically determine the widths of the HEIDI
scalar fields. In addition to the aforementioned decays into SM particles, two body
decays into two scalar modes are also included into the calculation.

\subsection{{\sc FeynRules} implementation}
Since the multi-scalar couplings are only of small phenomenological relevance
and at the same time slow down both {\sc FeynRules} and the matrix element generator
considerably (their number growing combinatorically with the number of modes),
we restrict the implementation to the SM sector of the model.
To this end, we insert Eq.~\eqref{equ:heidi-fieldexp} into the SM
Lagrangian and induce in this way the coupling between the HEIDI modes $\phi_k$
and the SM fermions and vector bosons, which is enough to study most of the LHC
phenomenology of this model. In addition, we restrict the implementation to
unitarity gauge. In the rest of this section we describe the
{\sc FeynRules} implementation of the model which is special in the sense that
the number of modes below the cutoff and thus the number of particles in the
model is variable.

Since the compact HEIDI model consists of the SM to which we add the HEIDI
sector, the {\sc FeynRules} implementation can be achieved in a natural way by
extending the already existing implementation of the SM in {\sc FeynRules},
contained in the file {\tt SM.fr} included in the {\sc FeynRules}
distribution. In this process\footnote{
As the SM Higgs field $h$ is no longer a mass eigenstate
in the model, it can be removed from the set of physical fields by adding
\texttt{Unphysical -> True} to its definition in \texttt{SM.fr}. However, this
step is strictly optional and no harm is done by omitting it.
}, \texttt{SM.fr} can remain unchanged, and the new model file only has to
encode the new parameters, fields and the modified SM Lagrangian.

Before the spectrum of the model can be calculated,
the Higgs vacuum expectation value $v$ and its pre-mixing mass $m_h$,
the bulk mass $m_b$, the
HEIDI coupling $g^2$ and the compactification scale $R^{-1}$ must be specified
(\emph{cf.} \secref{sec:heidi-spectrum}). In addition, for the actual implementation,
we need to truncate the infinite tower of Kaluza-Klein states to a finite number
by choosing a cutoff scale $\Lambda$. These six parameters, which must be known
\emph{before} any code for a matrix element generator can be emitted, have been
implemented as a set of flags which must be specified before the model can be
loaded successfully and which are listed in
\tabref{tab:heidi_flags}. It is important to realize that those
values are hardcoded in the generated model files. In particular, the
electroweak scale implied by the runtime input parameters \emph{must} match
the value of $v$ set by the aforementioned flags.
\begin{table}[!t]
\centerline{\begin{tabular}{|c|c|}
\hline
Flag  & Description \\\hline
\hline
{\hspace{5ex}\tt HEIDI\$v\hspace{5ex}} &  SM Higgs boson vev\\\hline
{\tt HEIDI\$mh} &  SM Higgs mass\\\hline
{\tt HEIDI\$cs} & compactification scale $R^{-1}$\\\hline
{\tt HEIDI\$mb} & bulk mass $m_b$\\\hline
{\tt HEIDI\$g2} &5D trilinear coupling squared\\\hline
{\tt HEIDI\$cutoff} & cutoff scale $\Lambda$\\\hline
{\tt HEIDI\$nmodes} & \parbox{0.7\textwidth}{\raggedright\strut Number of HEIDI
modes below the cutoff scale $\Lambda$ (determined automatically
if \texttt{HEIDI\$cutoff} is set)\strut}
\\\hline
\end{tabular}}
\caption{\label{tab:heidi_flags} The flags in the {\sc FeynRules}
implementation of the compact HEIDI model which must be set before the model can
be used.}
\end{table}

After setting the flags from \tabref{tab:heidi_flags}, the model such defined can be
loaded into {\sc FeynRules} and the {\sc Whizard} interface
can be invoked as described in \cite{Christensen:2010wz}.

\section{HEIDI scalars in the golden channel at the LHC}
\label{sec:golden}

In order to study the impact of HEIDI on one of the major Higgs search channels
at the LHC, we have performed simulations of scalar particle production
in gluon fusion with the subsequent decay into four leptons via two virtual $Z$
bosons, the so-called golden channel for Higgs discovery at the LHC.
As gluon fusion is a loop induced process with the dominant
contribution at leading order coming from the top quark triangle, we have added to the
Lagrangian the corresponding operator obtained by integrating out the
top quark (\emph{cf.}
\appref{app:heidi-hgg}).
\begin{figure}
\centerline{\begin{tabular}{c|cc}
Signal & \multicolumn{2}{c}{Background} \\\hline
& &  \\[-2ex]
\parbox{55mm}{\includegraphics{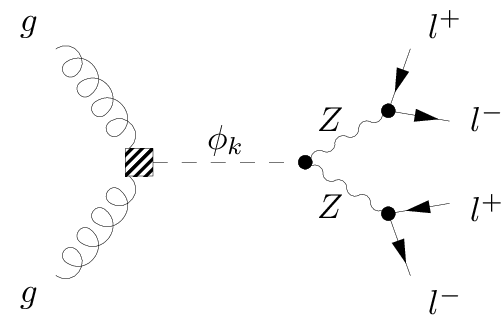}} &
\parbox{55mm}{\includegraphics{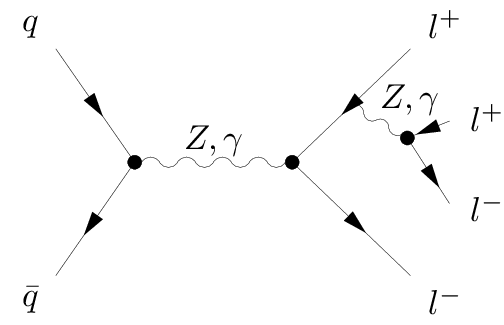}} &
\parbox{40mm}{\includegraphics{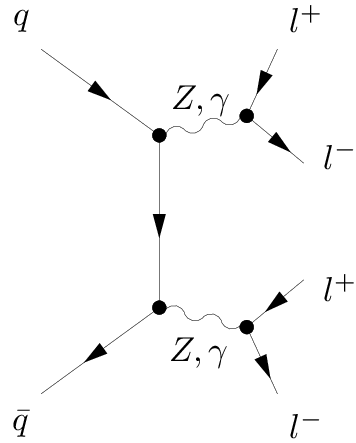}}
\end{tabular}}
\caption{Diagrams contributing to signal and background for scalar
  particle production in
$pp\rightarrow ZZ \rightarrow 4l$. The shaded box represents the effective gluon
fusion operator obtained from integrating out the top quark.}
\label{fig:heidi-ppzz-diags}
\end{figure}

At lowest order in the strong coupling, the signal in this process is
mediated by gluon initiated $s$-channel type diagrams
such as the one shown in \figref{fig:heidi-ppzz-diags} left, while the background
processes always have quarks in the initial state and consist of
diagrams such as
those in \figref{fig:heidi-ppzz-diags} right. Therefore, a significant part of
the background is contributed by processes with a valence and a sea quark in the
initial state. As those processes tend to be more strongly boosted when compared
to processes initiated by sea quarks or by gluons, it is possible to remove part
of the background by a cut on the total rapidity. We thus chose to apply a cut
of
\begin{equation*}
\left|y\right| \le 2
\end{equation*}
on the total rapidity of the final state.

Additional background can be removed by cutting the invariant mass of the
leptons to the $Z$ mass window. However, as we do not want to lose sensitivity to the
scalar modes which lie below the pair production threshold, we must allow for
lepton pairs with invariant mass significantly below the $Z$ mass and therefore
have enforced the cut
\begin{equation*}
\unit[10]{GeV} \le m_{ll} \le \unit[100]{GeV}
\end{equation*}
on the invariant mass of the lepton pairs. If the final state consists of four
leptons of the same generation, we demanded that at least one of the two possible
combinations satisfies the cut. In order to account for detector acceptance
and for the fact that we require four separately resolved leptons, we have applied
a $p_T$ cut as well as cuts on the angular separation and on the polar angle
of the leptons
\begin{equation*}
p_T > \unit[5]{GeV} \qquad,\qquad \left|\cos\theta_{ll}\right| \le 0.99
   \qquad,\qquad \left|\cos\theta_l\right| \le 0.99 \ .
\end{equation*}

\begin{figure}
\centerline{\begin{tabular}{ccc}
\includegraphics[angle=270,width=0.48\textwidth]{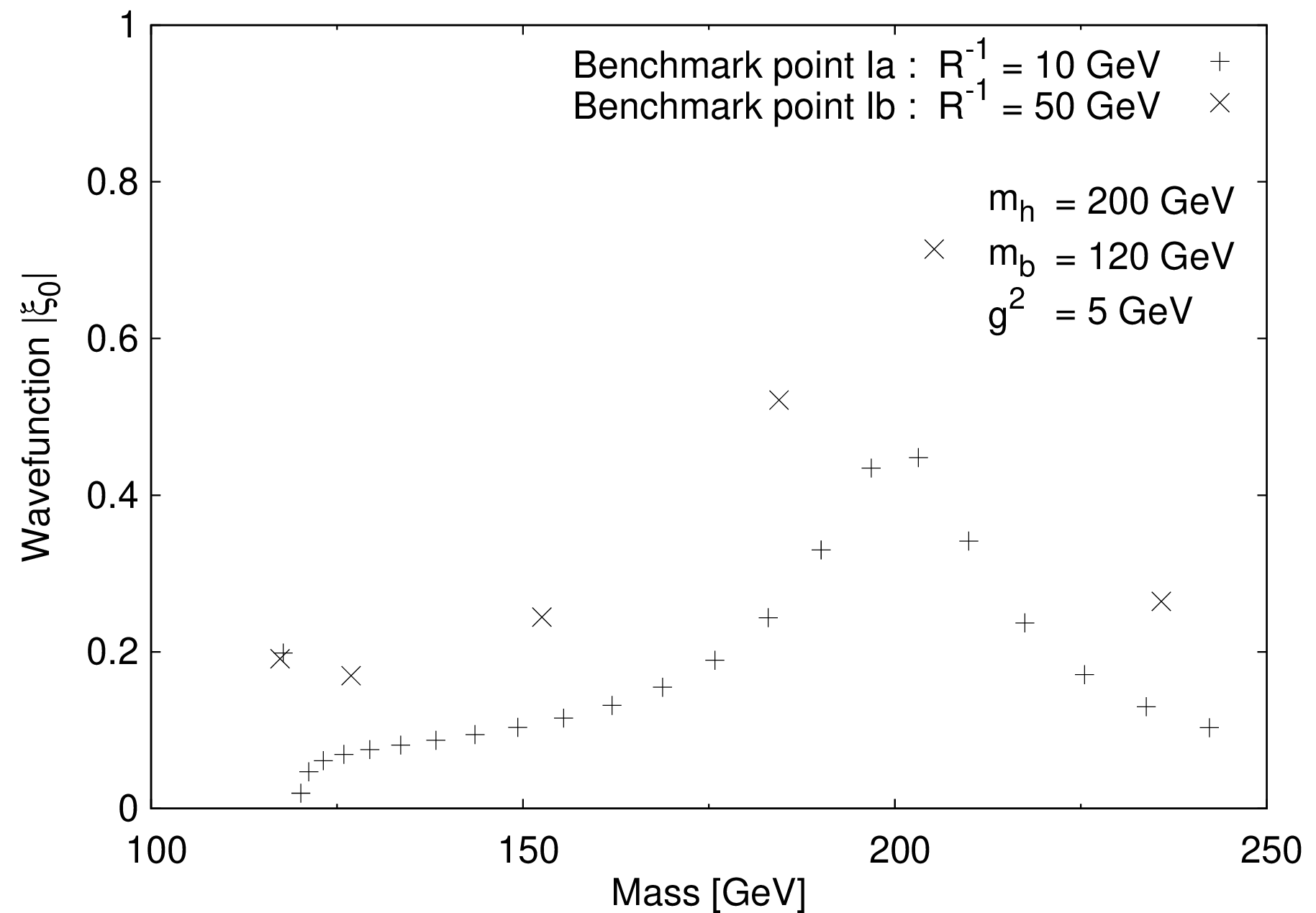} &
\hspace{2ex} &
\includegraphics[angle=270,width=0.48\textwidth]{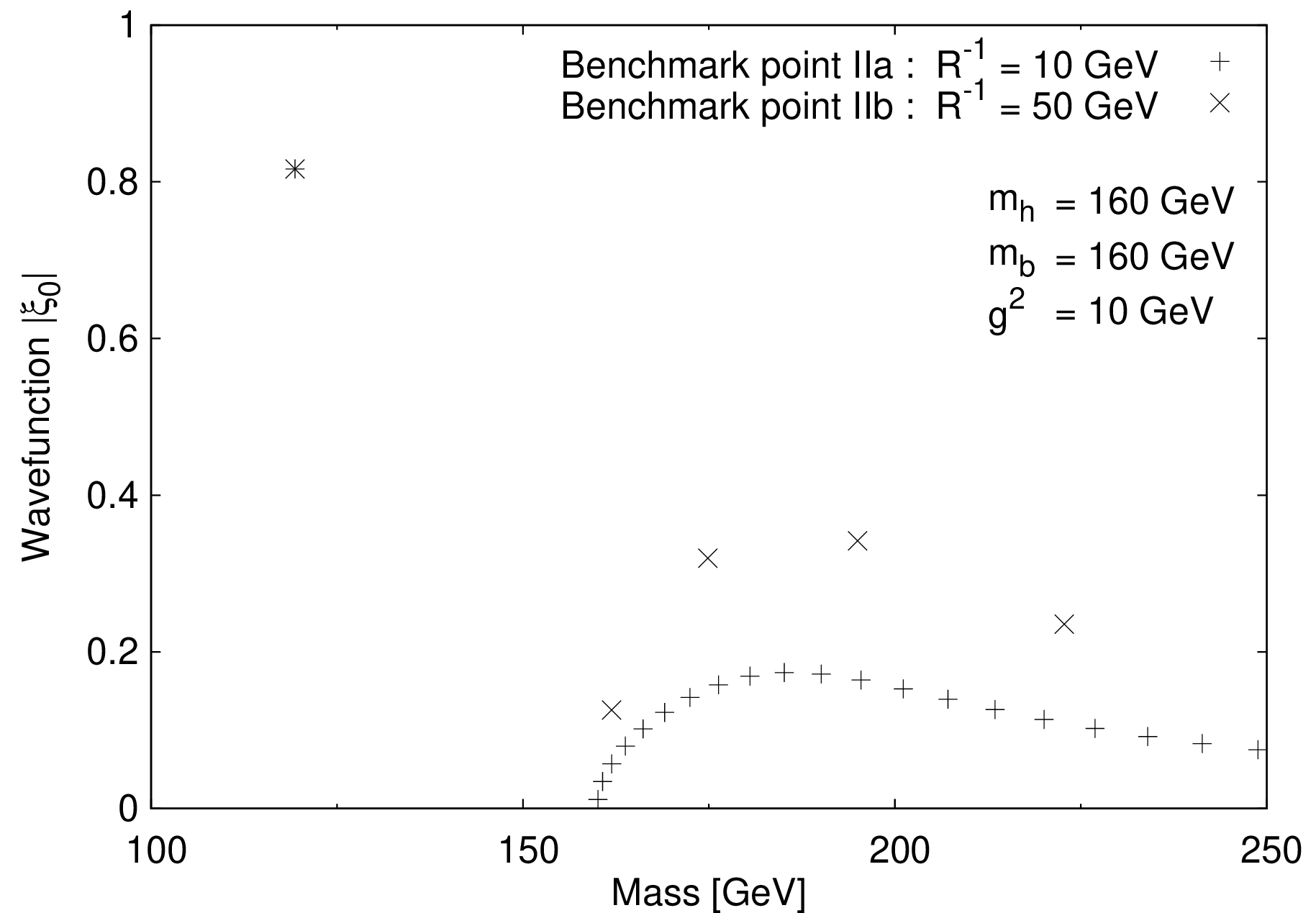}
\end{tabular}}
\caption{The scalar field spectra at the four benchmark points at which simulations
have been performed.}
\label{fig:heidi-parpoints}
\end{figure}

We have performed simulations at four different benchmark points in parameter
space which are shown in \figref{fig:heidi-parpoints}. The parameter sets
Ia and Ib differ only by
the compactification scale and have $m_h = \unit[200]{GeV}$ and
$m_b=\unit[120]{GeV}$. As discussed in \secref{sec:heidi-parspace}, this implies
that the would-be SM Higgs boson corresponds to a bump in the would-be 
continuum
around $m_h$ as clearly visible in the spectra shown in
\figref{fig:heidi-parpoints} left. At point Ia, the compactification scale is
$R^{-1}=\unit[10]{GeV}$ with more than 20 modes lying below $\unit[250]{GeV}$,
while we only have 6 such modes at Ib with $R^{-1}=\unit[50]{GeV}$,
their couplings to the SM particles being enhanced in exchange.

At IIa and IIb, we set $m_h = m_b = \unit[160]{GeV}$. As discussed in
\secref{sec:heidi-parspace} and clearly visible from
\figref{fig:heidi-parpoints} on the right, the spectrum is very different in this case,
with the lowest mode corresponding to the would-be Higgs boson, and the couplings of
the would-be continuum to the SM being strongly suppressed. Again, we chose
$R^{-1}=\unit[10]{GeV}$ for IIa and $R^{-1}=\unit[50]{GeV}$ for IIb.

For the actual simulation, we took the first two quark generations and gluons
into account as potential initial particles, simulated
the full $2\rightarrow 4$ processes
using the CTEQ6L \cite{Pumplin:2002vw} series of parton distribution functions and
generated unweighted events for an integrated luminosity of $\unit[100]{fb^{-1}}$
at $\sqrt{s}=\unit[14]{TeV}$. As the background is the same for all four
benchmark points, we generated a single set of background events, while a fresh
set of signal events was generated for each point.

\begin{figure}
\centerline{\begin{tabular}{ccc}
\includegraphics[angle=270,width=0.48\textwidth]{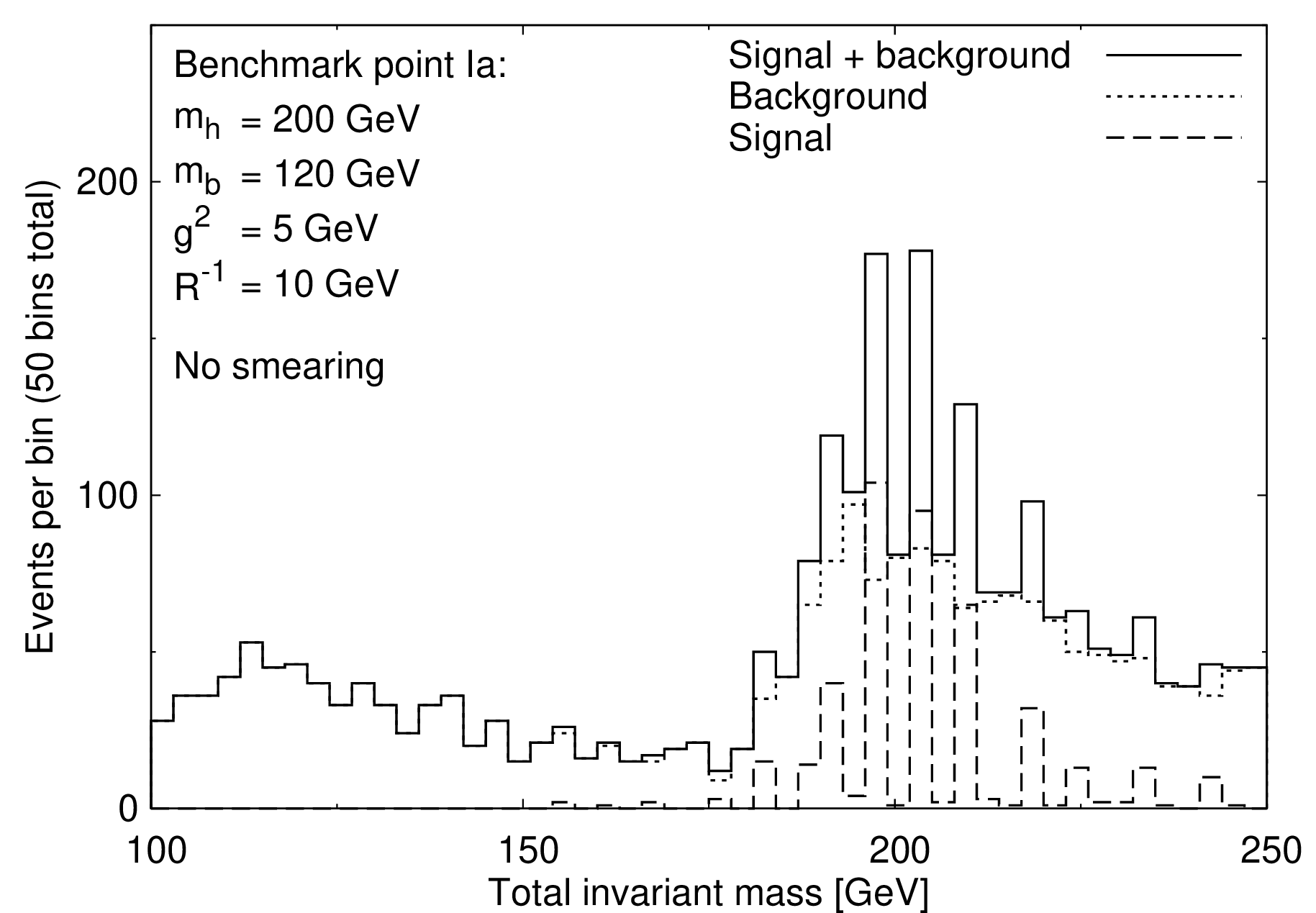} &
\hspace{2ex} &
\includegraphics[angle=270,width=0.48\textwidth]{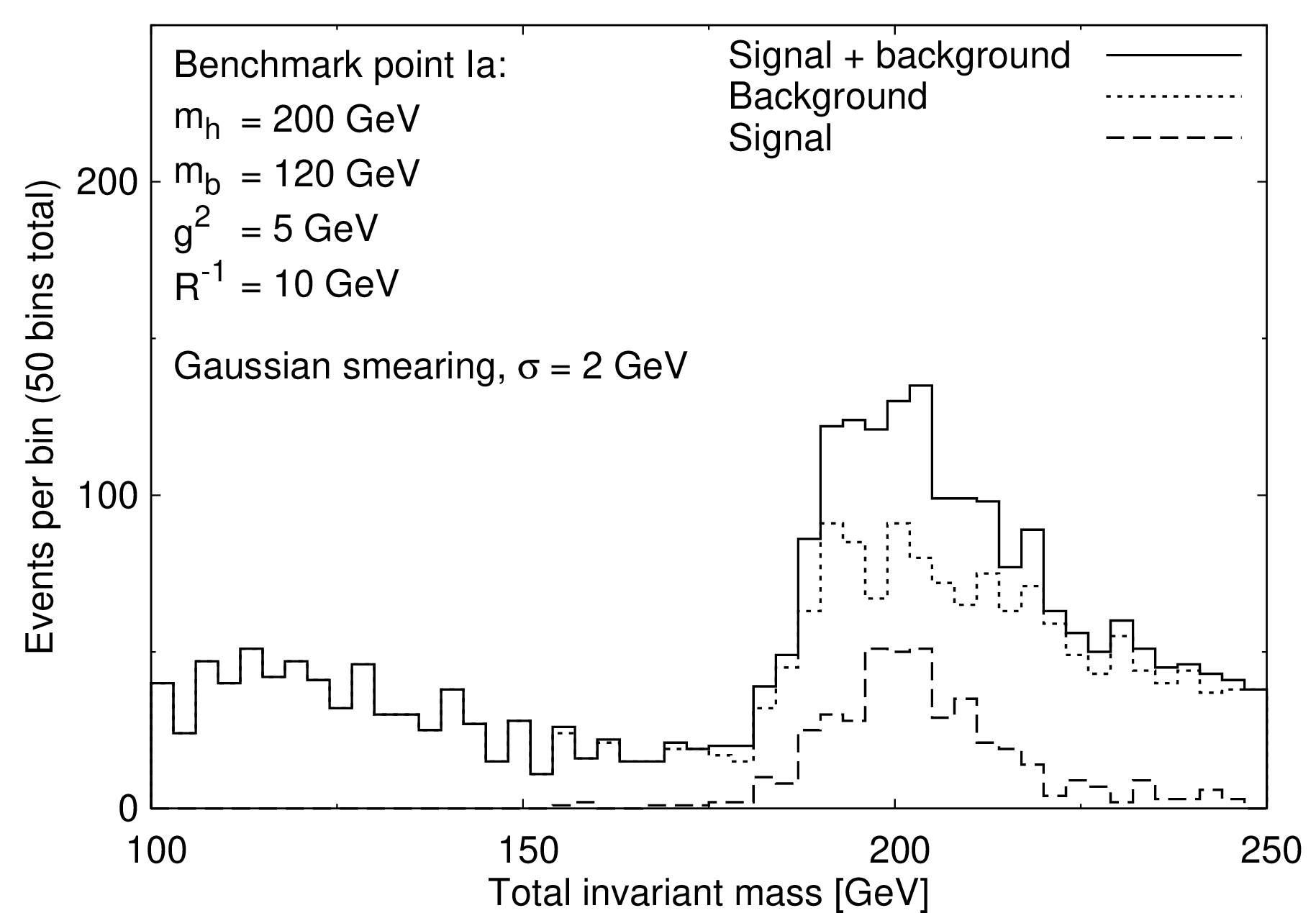}
\\
\includegraphics[angle=270,width=0.48\textwidth]{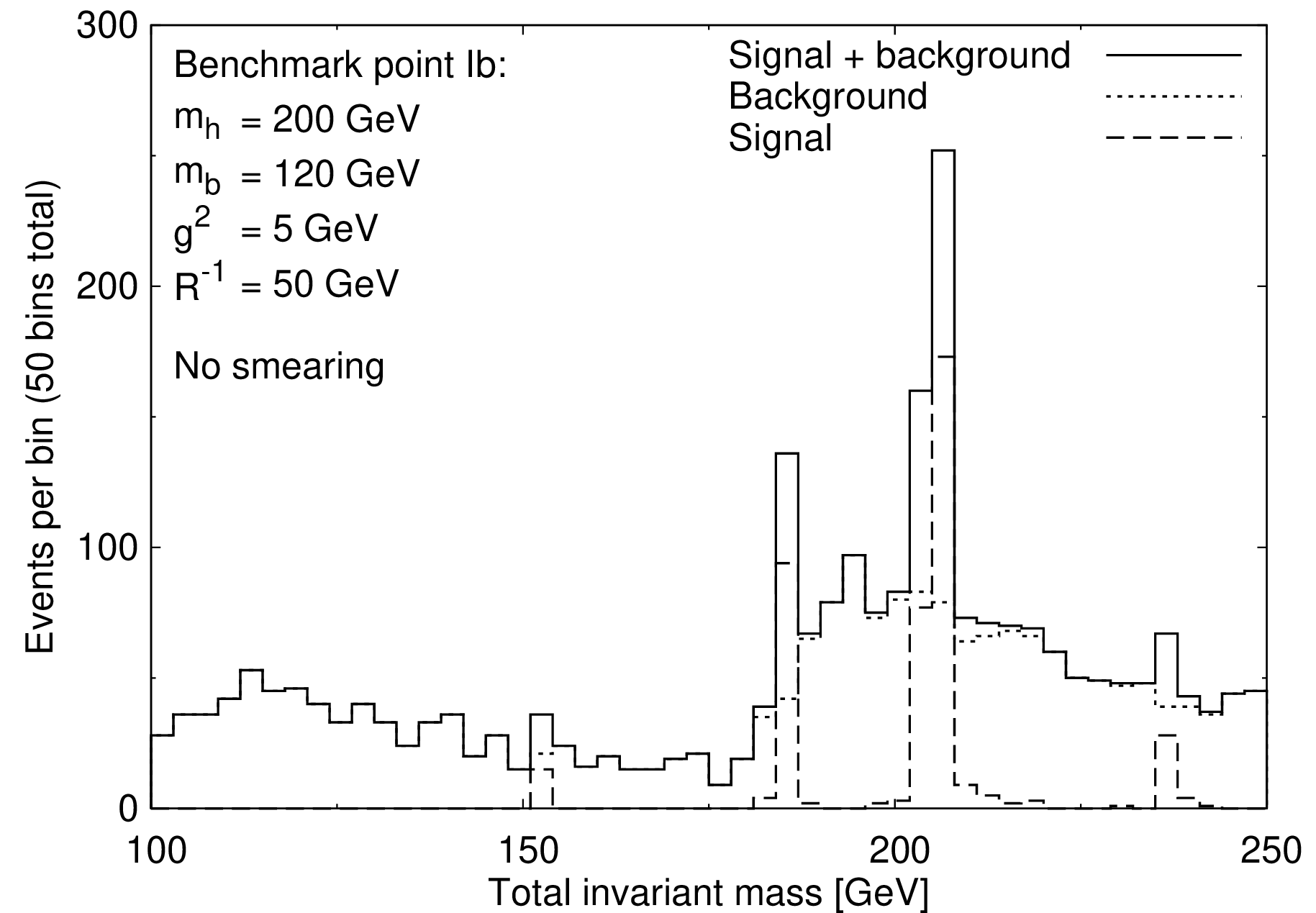} &
\hspace{2ex} &
\includegraphics[angle=270,width=0.48\textwidth]{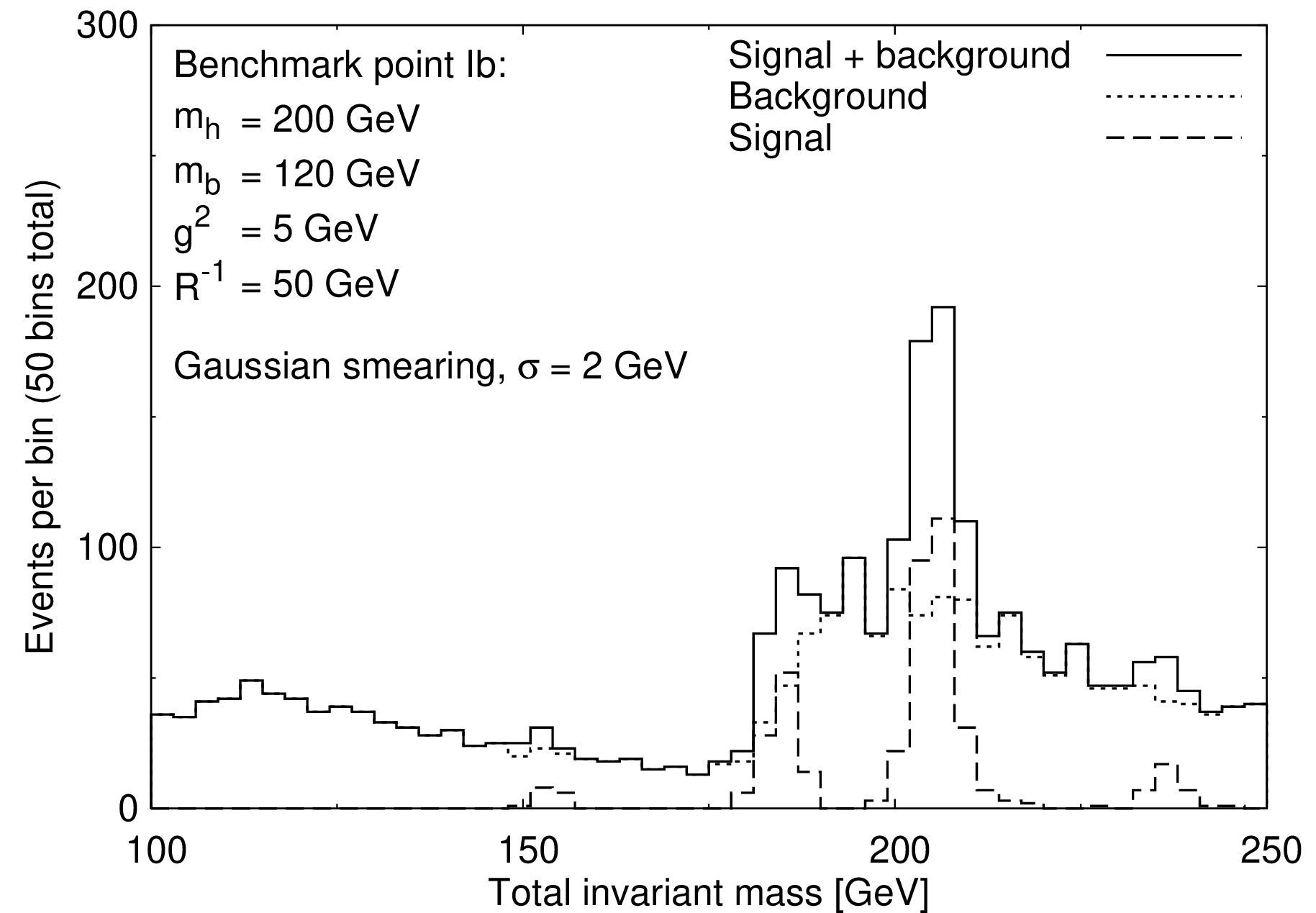}
\end{tabular}}
\caption{Total invariant mass distributions in $pp\rightarrow 4l$ at
benchmark points Ia and Ib. For the distributions in the right column, a Gaussian
smearing with $\sigma=\unit[2]{GeV}$ has been applied to the invariant mass
in order to simulate measurement errors.}
\label{fig:heidi-hist-I}
\end{figure}

The top left panel of \figref{fig:heidi-hist-I} shows the
resulting invariant mass distribution
for benchmark point Ia. Despite the fact that the scalar field tower already starts at
$\unit[120]{GeV}$, the actually visible resonances only start at the $Z$ pair
production threshold, demonstrating that the SM couplings of those modes are much
too small to win over the suppression coming from the off-shell $Z$ propagators.
Above the threshold, the comblike structure coming from the scalar resonances
is clearly visible not only in the actual signal, but persists also when the
background is added.

The same distributions are shown in the top right panel of
\figref{fig:heidi-hist-I}, but now with a
Gaussian smearing with standard deviation $\sigma=\unit[2]{GeV}$ applied to the
invariant mass in order to simulate a measurement error. Although the width of
the Gaussians is still significantly smaller than the average mode spacing of
about $\unit[10]{GeV}$, the histogram shows that this is already enough to
destroy the comblike structure of the would-be continuum. What remains is a broad
excess above the background which peaks around $m_h$ and resembles a single,
very broad resonance.

The bottom panels of \figref{fig:heidi-hist-I} show the unsmeared (left) and
smeared (right) distributions for benchmark point Ib. As detailed above, the
only difference with point Ia is the larger spacing of the individual modes and the
therefore enhanced couplings to the SM. Before the smearing is applied, four modes
are visible in the signal (one of them below the $Z$ pair threshold with a mass of
$\approx\unit[150]{GeV}$), and the three heaviest of them also stick out
above the background. While the smearing distorts the shape of the peaks
and makes them less pronounced, the smeared peaks are still visible
above the background.

\begin{figure}
\centerline{\begin{tabular}{ccc}
\includegraphics[angle=270,width=0.48\textwidth]{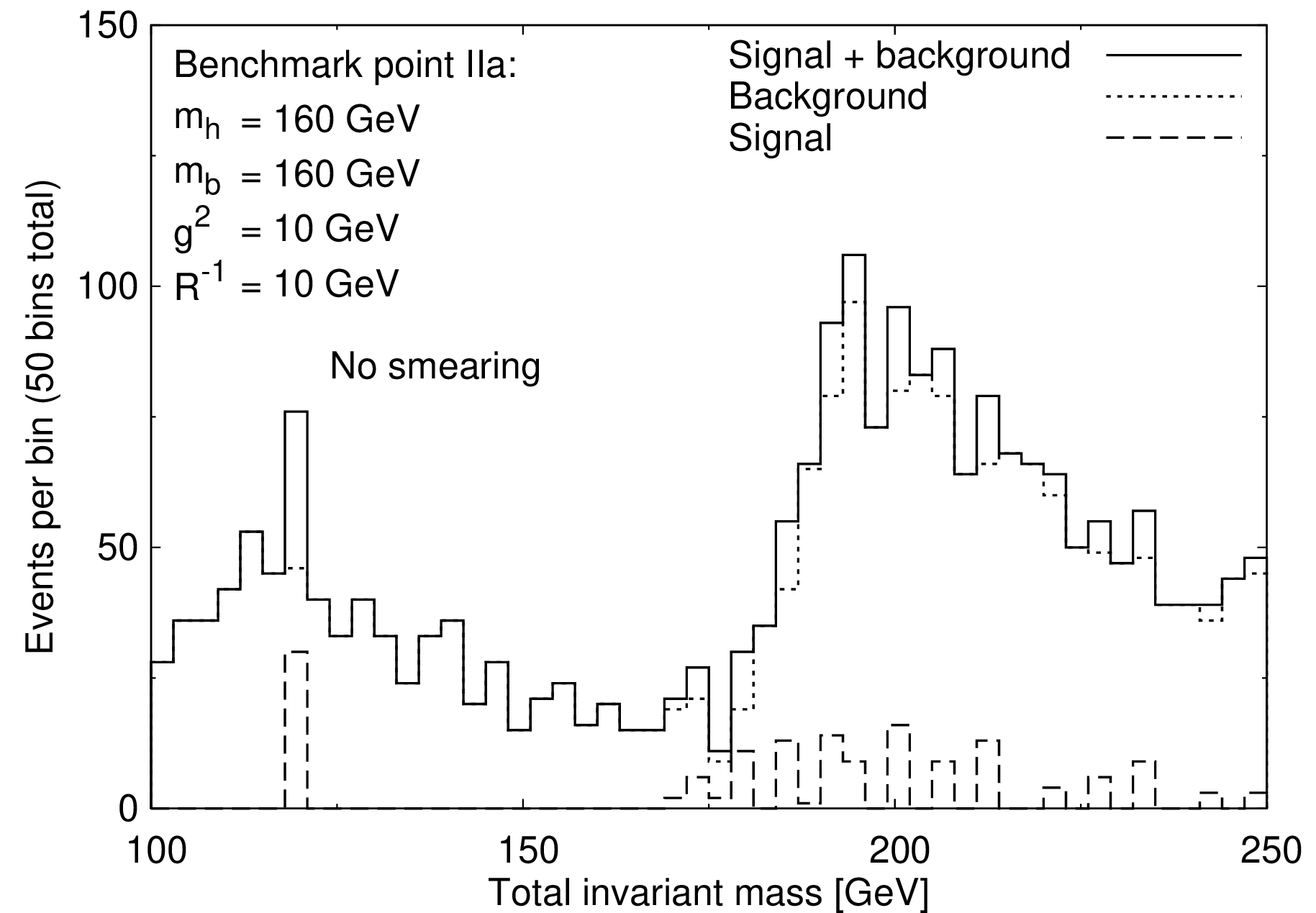} &
\hspace{2ex} &
\includegraphics[angle=270,width=0.48\textwidth]{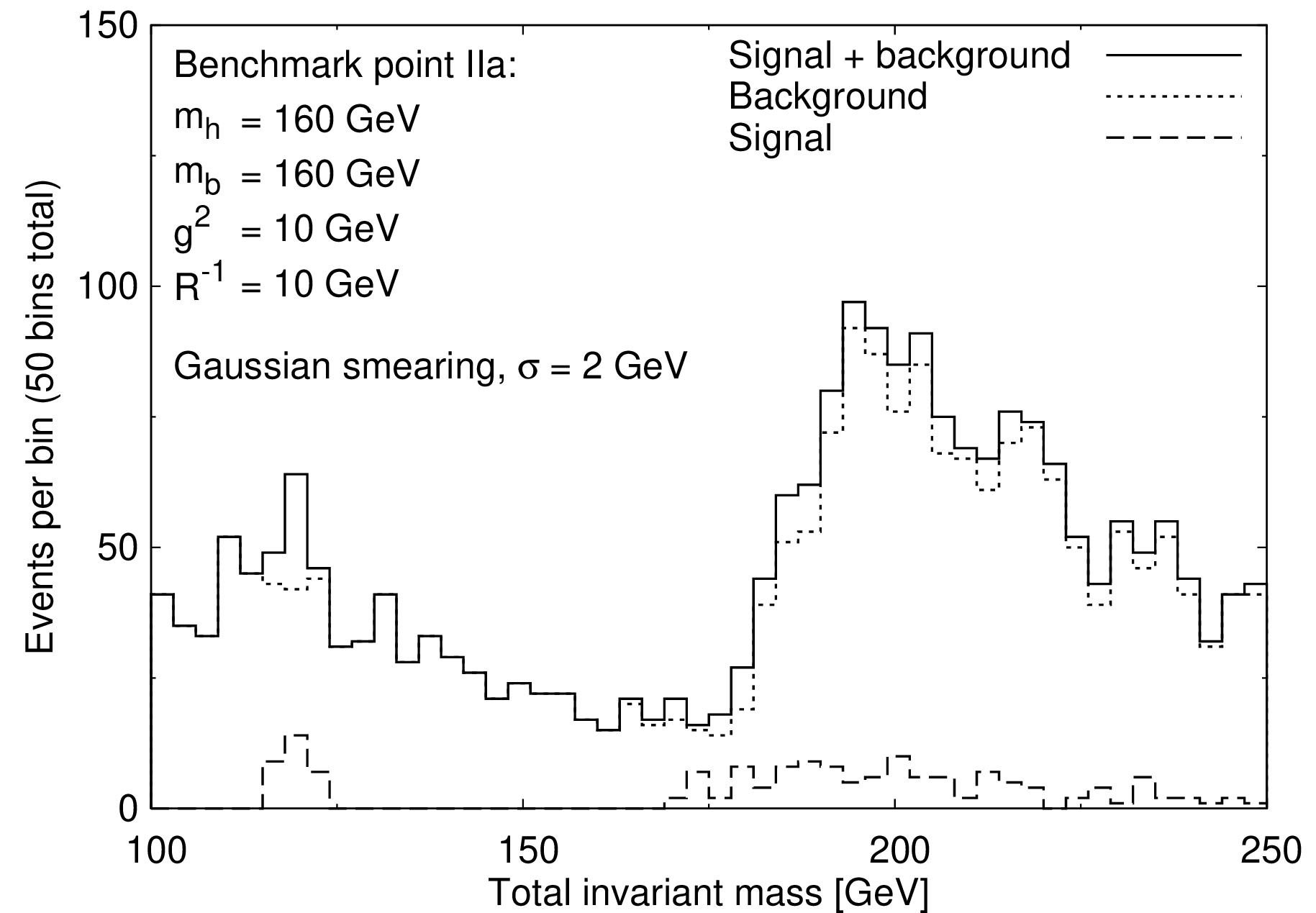}
\\
\includegraphics[angle=270,width=0.48\textwidth]{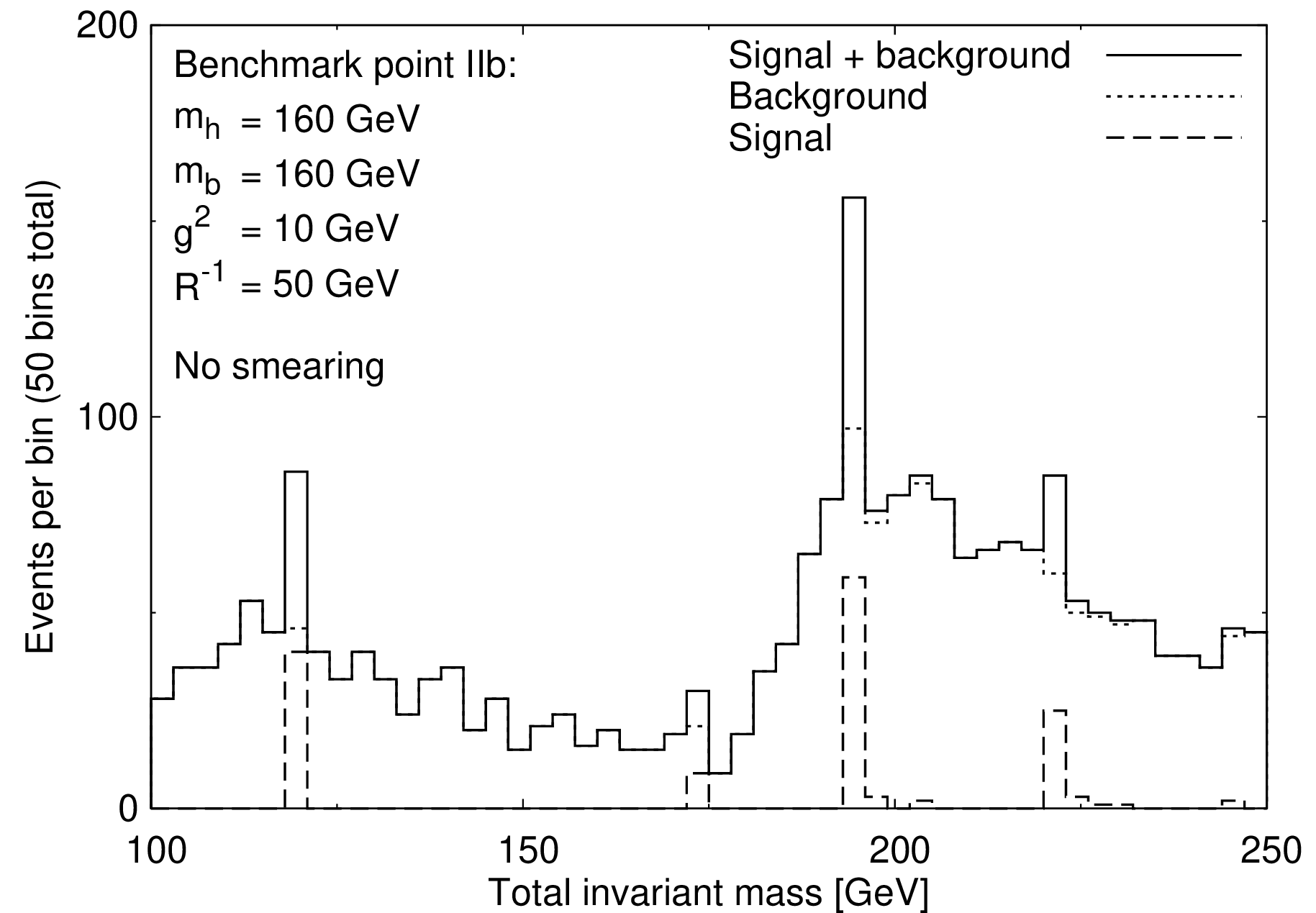} &
\hspace{2ex} &
\includegraphics[angle=270,width=0.48\textwidth]{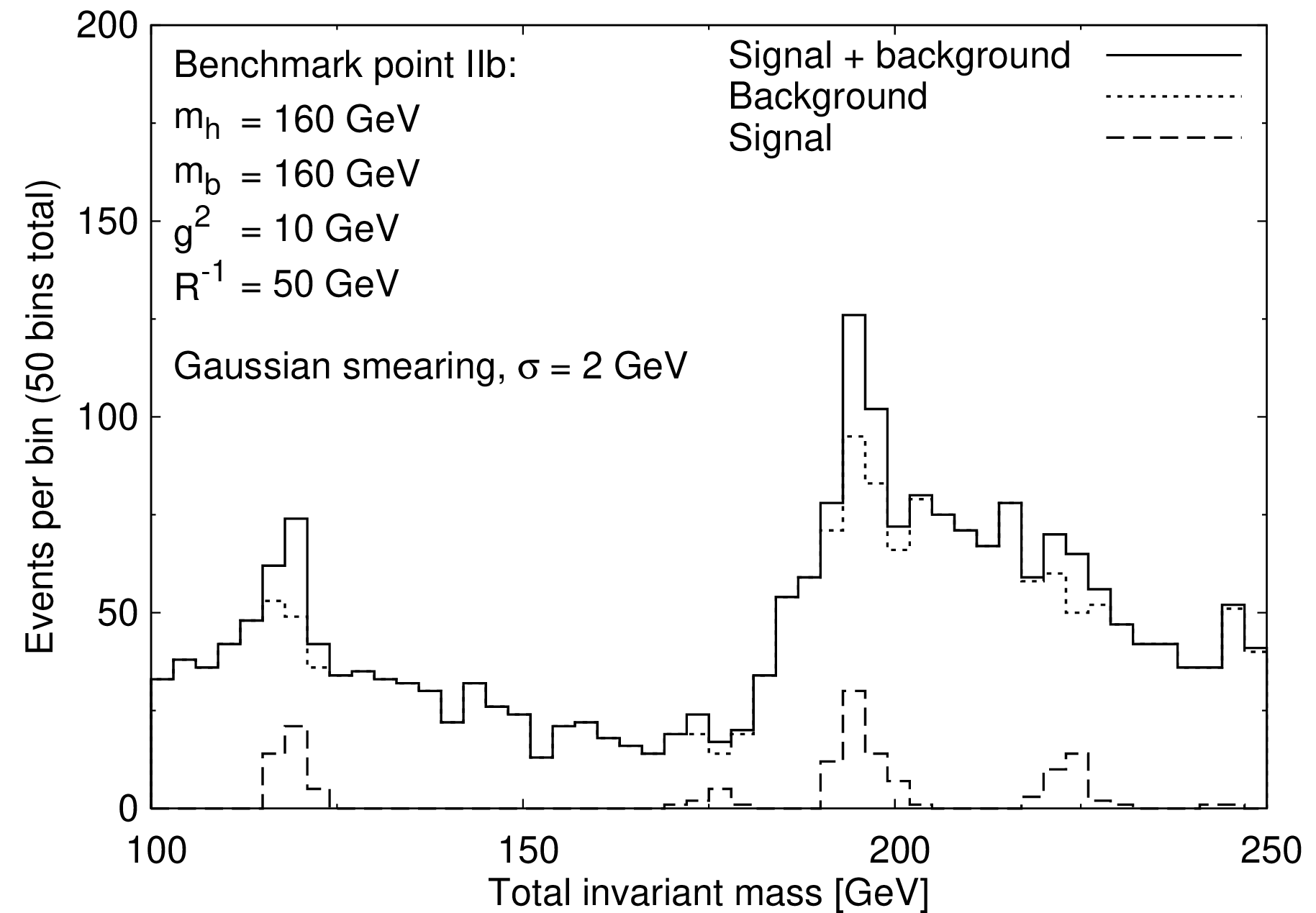}
\end{tabular}}
\caption{Like \figref{fig:heidi-hist-I}, but for the benchmark points IIa and
IIb.}
\label{fig:heidi-hist-II-IV}
\end{figure}
In the same way, the invariant mass distributions for benchmark points IIa and
IIb are shown in \figref{fig:heidi-hist-II-IV}, the top panels again showing the
case with the smaller mode spacing. The different structure of the spectrum
discussed above also manifests itself in the distributions. The lowest mode,
which again lies at about $\unit[120]{GeV}$, has nearly SM-like couplings in this
scenario and is therefore a prominent feature in the signal distribution.
Despite lying far below the $Z$ pair threshold, the corresponding peak is
more than comparable in size with the peaks from the would-be continuum.
After adding the background, the lowest mode
still sticks out and remains visible, while the comblike structure of the other
modes is masked by the background fluctuations with only a diffuse excess being
observable. Applying the smearing degenerates the peak of the lowest mode and
homogenizes the (small) excess over the background.

The distributions for benchmark point IIb again differ by the mode count. While
the appearance of the lowest mode is virtually identical to scenario IIa (as is
clear from \figref{fig:heidi-parpoints}), the would-be continuum now exhibits
two moderate peaks in the signal, both of which survive when the background is
added. After the smearing is added, at least one of those two peaks remains
visible over the background together with the lowest mode.

Our simulations show that the model could, depending on
the scenario realized, appear in very different guises, ranging from a Higgs-like
peak with reduced magnitude and a diffuse continuum (IIa), a series of sharp
resonances (IIb and Ib) or a single, very broad excess (Ia). At least for Ia /
Ib, the signal is clearly visible over the background and a discovery might
be possible in this channel. For IIb and especially IIa, the situation looks more
pessimistic; especially in IIa where the diffuse shape of the
signal might get lost in the background. However, while being
the first detailed exploration of a HEIDI-like scalar sector at the LHC, our
simulations are still too simplistic to make realistic efficiency estimates for a
discovery. To this end, a more detailed study of the background and of higher order
corrections is required as well as the inclusion of other potential discovery
channels like gauge boson fusion and ``HEIDI strahlung''. We
postpone this to a future publication.

\section{Conclusions}

In this paper, we have presented a first study of the LHC phenomenology of an
extended scalar sector built from a compact extra dimension, where the
SM fields are living
on a brane and only a single singlet scalar field is allowed to
propagate in the bulk. This is a representative of a class of models known as
higher-dimensional (``HEIDI'') Higgs models.

Despite the presence of just a single higher-dimensional scalar 
singlet, the physics and phenomenology of
this model is astonishingly rich. The main effect comes from the
Kaluza-Klein tower of scalar resonances mixing with the SM Higgs
scalar field. As the extra-dimensional scalar fields couple to SM
particles through this mixing, the SM Higgs boson gets
diluted to a tower of scalar fields of ascending mass whose couplings to the SM 
particles are reduced significantly compared to the SM Higgs boson.

We concentrated on the so-called ``golden channel'' for a not too
light Higgs boson decaying predominantly into two $Z$ bosons, the latter 
decaying into four leptons, and restricted ourselves to the gluon-fusion production
channel. Our main results in \secref{sec:golden} show a
variety of different shapes which such a model could produce,
depending on the point in parameter space under consideration. These shapes 
range from a simple
reduction of a SM Higgs signal with a diffuse continuum tail to a series
of Kaluza-Klein peaks to a very broad continuous bump,
showing the phenomenological richness of these otherwise simple extensions 
of the SM. While this phenomenological variety is a generic property of the
model which derives from its spectrum, it would certainly be interesting to see
how the other classic Higgs production channels perform for this type
of model. We conclude by remarking that even a rather trivial deformation of the
SM in the scalar sector can have the potential to jeopardize the common assumption
that the LHC cannot fail in its mission to discover either the Higgs boson or new
physics.


\section*{Acknowledgments}
\label{sec:conclusions}

CS has been supported by the Deutsche Forschungsgemeinschaft through
the Research Training Groups GRK\,1147
\textit{Theoretical Astrophysics and Particle Physics} and  GRK\,1102
\textit{Physics of Hadron Accelerators}. JRR has been partially supported
by the Ministry of Science and Culture (MWK) of the German state
Baden-W\"urttemberg. BF acknowledges support by the Theory-LHC 
France-initiative of the CNRS/IN2P3. NDC was supported by the US
National Science Foundation under grants PHY-0354226 and PHY-0705682.
We are also indebted to J. van der Bij for many instructive
discussions on the nature of HEIDI-type 
models.


\begin{appendix}


\section{Diagonalizing compact HEIDI}
\label{app:heidi-notes}

\subsection{Vacuum expectation values}
\label{app:heidi-vev}

From Eq.~\eqref{equ:heidi-s5d} and Eq.~\eqref{equ:heidi-s4d}, we can read off the
classical potential for the vacuum expectation values of the
fields
Eq.~\eqref{equ:heidi-shifts},
\begin{equation}
V(v, w) = \frac{1}{4}\lambda v^4 - \frac{1}{2}\mu^2v^2 +
   \sum_{k=0}^\infty \underbrace{
      \left( \frac{1}{2}m_k^2w_k^2 -
         \frac{g}{N_k \sqrt{R}}v^2 w_k \right)}_{
      V_k(v, w_k)}  \; .
\label{equ:heidi-potential}
\end{equation}
For any fixed value of $v$, Eq.~\eqref{equ:heidi-potential} decomposes into a
constant term plus a sum of second degree polynomials $V_k$, each depending
only on a single $w_k$. If we have $m_b>0$, then each of the $V_k$ is bounded
from below, and the the minima are given by
\begin{equation}
w_k = \frac{g}{N_k\sqrt{R}m_k^2}v^2 \; .
\label{equ:heidi-pot-w}
\end{equation}
Therefore, the potential $V$ is always bounded from below as a function of the
$w_k$ for fixed $v$ and $m_b > 0$, and inserting Eq.~\eqref{equ:heidi-pot-w} into
Eq.~\eqref{equ:heidi-potential} gives the minimum
\begin{equation}
V_0(v) = \frac{1}{4}\left(\lambda - \alpha\right)v^4 - \frac{1}{2}\mu^2v^2 \; .
\label{equ:heidi-pot-eff}
\end{equation}
With the help of
\begin{equation}
\sum_{k=0}^\infty\frac{1}{1 + \frac{k^2}{x^2}} = \frac{1}{2} + \frac{\pi x}{2}
   \coth\left(\pi x\right) \; ,
\label{equ:heidi-mastersum}
\end{equation}
$\alpha$ in Eq.~\eqref{equ:heidi-pot-eff} can be written as
\begin{equation}
\alpha = \frac{g^2}{R} \sum_{k=0}^\infty \frac{1}{N_k^2 m_k^2} =
   \frac{g^2}{m_b}\coth\left(R\pi m_b\right) \; .
\label{equ:heidi-alpha}
\end{equation}

In order to derive the conditions under which Eq.~\eqref{equ:heidi-potential} is
bounded from below as a function of both $v$ and the $w_k$, it is sufficient
to examine the asymptotic behavior of
Eq.~\eqref{equ:heidi-pot-eff} which is determined by the sign of the quartic term.
As $\alpha$ can be easily seen to be positive definite from
Eq.~\eqref{equ:heidi-alpha}, we end up with four different scenarios:\\
\begin{enumerate}
\item $\mu^2 > 0 \quad,\quad \lambda > \alpha$:\\
$V$ is bounded from below and has two minima at
\begin{equation}
v = \pm\frac{\mu}{\sqrt{\lambda - \alpha}} \quad,\quad
w_k=\frac{g}{N_k\sqrt{R}m_k^2}v^2 \; .
\label{equ:heidi-vev-minima}
\end{equation}
The point $v=w_k=0$ is unstable.
\item $\mu^2 > 0 \quad,\quad \lambda \le \alpha$:\\
The potential is not bounded.
\item $\mu^2 \le 0 \quad,\quad \lambda \ge \alpha$:\\
$V$ is bounded from below and has a single minimum at $v=w_k=0$
\item $\mu^2 \le 0 \quad,\quad \lambda < \alpha$:\\
The potential is not bounded.
\end{enumerate}
\mbox{}\vspace{1ex}

Therefore, we need $\mu^2 > 0$ and $\lambda > \alpha$ in order to have a stable
theory with spontanous symmetry breaking. The two solutions
Eq.~\eqref{equ:heidi-vev-minima} are equivalent and only differ by a field
redefinition $h \rightarrow -h$ as can be seen from Eq.~\eqref{equ:heidi-lphys},
and we therefore are free to make the choice Eq.~\eqref{equ:heidi-vevs}.

\subsection{Mass eigenstates and wave functions}
\label{app:heidi-wvs}

Arranging the fields $h$ and $\omega_k$ into the vector
\begin{equation*}
\Upsilon = \left(h, \omega_0, \omega_1, \dots\right)^T
\end{equation*}
and introducing the mass matrix $M$
\begin{equation}
\left(M\Upsilon\right)_k = \begin{cases}
m_h^2 h + \sum_{l=0}^\infty \rho_l \omega_l\quad
   & \text{for } k=0 \\
\rho_{k-1} h + m_{k-1}^2 \omega_{k-1} & \text{for } k > 0
\end{cases}
\end{equation}
with the abbreviation
\begin{equation*}
\rho_k = \frac{2gv}{N_k\sqrt{R}} \ ,
\end{equation*}
the mass and mixing terms in Eq.~\eqref{equ:heidi-lphys} can be written as
\begin{equation*}
-\frac{1}{2}m_h^2 h^2 - \sum_{k=0}^\infty\left(
   \frac{1}{2}m_k^2 \omega_k^2 - \frac{2gv}{N_k\sqrt{R}}h\omega_k\right) =
-\frac{1}{2}\Upsilon^T M \Upsilon \ .
\end{equation*}
In order to find the eigenvalues $\lambda_k^2$ to the eigenvectors\footnote{Note
that this definition of the $\xi^k_i$ is equivalent to that given in
Eq.~\eqref{equ:heidi-xidef}.}
\begin{equation}
\xi^k = \left(\xi^k_0, \xi^k_1, \dots\right)^T \qquad,\qquad
M\xi^k = \lambda^2_k\xi^k
\label{equ:heidi-evcond} \ ,
\end{equation}
we rewrite the eigenvalue condition of Eq.~\eqref{equ:heidi-evcond} in terms of the
components $\xi^k_i$
\begin{equation}
\lambda^2_k\xi^k_0 = m_h^2 \xi^k_0 + \sum_{i=0}^\infty \rho_i \xi^k_{i+1}
\label{equ:heidi-diag-1} \ ,
\end{equation}
\begin{equation}
\lambda^2_k\xi^k_i = \rho_{i-1} \xi^k_0 + m_{i-1}^2 \xi^k_i
\qquad\text{for}\qquad i > 0
\label{equ:heidi-diag-2} \ . 
\end{equation}
Since assuming $\xi^k_0=0$ and $\xi^k_i\neq0$ for some $i>0$ leads to a contradiction, we can therefore solve
Eq.~\eqref{equ:heidi-diag-2} in terms of $\xi^k_0$,
\begin{equation}
\xi^k_i = \xi^k_0\frac{\rho_{i-1}}{\lambda_k^2 - m_{i-1}^2}
\label{equ:heidi-evect1} \ . 
\end{equation}
Upon inserting Eq.~\eqref{equ:heidi-evect1} into Eq.~\eqref{equ:heidi-diag-1}, $\xi^k_0$
drops out and application of Eq.~\eqref{equ:heidi-mastersum} finally leads to the
result Eq.~\eqref{equ:heidi-mtrans}.
In order to obtain the normalized eigenvectors $\xi^k$, we rewrite the
normalization condition using Eq.~\eqref{equ:heidi-evect1}
\begin{equation}
1 = \sum_{i=0}^\infty \left(\xi_i^k\right)^2 = \left(\xi_0^k\right)^2
\left(1 + \sum_{i=0}^\infty \left(\frac{\rho_i}{\lambda_k^2 -
m_i^2}\right)^2\right)
\label{equ:heidi-norm} \ .
\end{equation}
The sums appearing in Eq.~\eqref{equ:heidi-norm} can be performed analytically,
and solving this equation for $\xi^k_0$, one obtains,
\begin{equation}
\left(\xi_0^k\right)^{-2} = 1 + \frac{g^2v^2R\pi}{\lambda_k^2 - m_b^2} +
\frac{1}{2}\,\frac{\lambda_k^2 - m_h^2}{\lambda_k^2 - m_b^2} +
\frac{R\pi}{4g^2v^2}\left(\lambda_k^2 - m_h^2\right)^2
\label{equ:heidi-evect2} \ .
\end{equation}
The other components of the eigenvectors can be derived by inserting
Eq.~\eqref{equ:heidi-evect2} into Eq.~\eqref{equ:heidi-evect1}.

\subsection{Feynman rules}
\label{app:heidi-fr}

As the model differs from the SM in the scalar sector only, all
couplings involving fermions and gauge bosons are unchanged and
need not be repeated here. To obtain the Feynman rules
involving the scalar fields, we have to take the full Lagrangian of the model
(which consists of the SM Lagrangian together with the scalar sector
Eq.~\eqref{equ:heidi-lphys}) and express $h$ and the $\omega_k$ through the mass
eigenstates $\phi_k$ by application of Eq.~\eqref{equ:heidi-fieldexp}.

For the couplings between the $\phi_k$ and the SM vectors and fermions, the
resulting Feynman rules are trivially obtained by taking the SM rules and
replacing the Higgs legs with the $\phi_k$, multiplying with a factor $\xi^k_0$
for every scalar leg. The results are Feynman rules which look like\\[2ex]
\begin{tabular}{rcl}
\parbox{0.3\textwidth}{\includegraphics{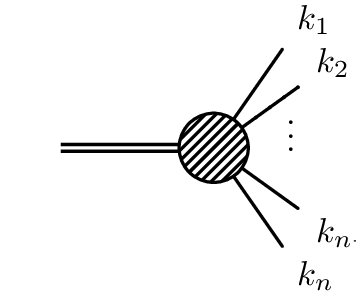}}
&
\parbox{0.1\textwidth}{$\displaystyle =$}
&
\parbox{0.6\textwidth}{$\displaystyle \Gamma\;\xi^{k_1}_0\dots\xi^{k_n}_0 \ ,$}
\end{tabular}\mbox{}\\[2ex]
where the double line represents all vector and fermion lines, and
$\Gamma$ is the SM vertex factor (including the color and Lorentz
structures).

The trilinear scalar couplings are more complicated: the piece of
Eq.~\eqref{equ:heidi-lphys} which encodes them is
\begin{equation*}
\mathcal{L}_3 = - \lambda vh^3 + \sum_{k=0}^\infty
   \frac{g}{N_k\sqrt{R}}h^2\omega_k \ . 
\end{equation*}
To obtain the corresponding Feynman rule in a compact form, we insert the
physical fields via Eq.~\eqref{equ:heidi-fieldexp}, sum the series using
Eq.~\eqref{equ:heidi-evect1} and Eq.~\eqref{equ:heidi-mastersum},
apply Eq.~\eqref{equ:heidi-mtrans} and replace $\lambda$
using Eq.~\eqref{equ:heidi-mh}. We eventually derive\\[2ex]
\begin{tabular}{rcl}
\parbox{0.3\textwidth}{\includegraphics{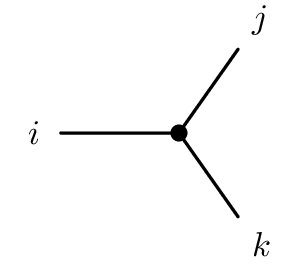}}
&
\parbox{0.1\textwidth}{$\displaystyle =$}
&
\parbox{0.6\textwidth}{$\displaystyle
\frac{i}{v}\xi^i_0\xi^j_0\xi^k_0 \left( \lambda_i^2 + \lambda_j^2 +
\lambda_k^2 - 6m_h^2\right)
$\ .}  
\end{tabular}\mbox{}\\[2ex]

The quartic couplings are considerably simpler to obtain since it is sufficient
to insert Eq.~\eqref{equ:heidi-fieldexp} into the $h^4$ term of
Eq.~\eqref{equ:heidi-lphys}. We directly get\\[2ex]
\begin{tabular}{rcl}
\parbox{0.3\textwidth}{\includegraphics{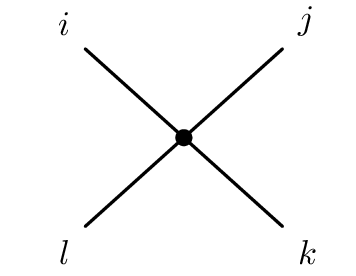}}
&
\parbox{0.1\textwidth}{$\displaystyle =$}
&
\parbox{0.6\textwidth}{$\displaystyle
-i3!\lambda\;\xi^i_0\xi^j_0\xi^k_0\xi^l_0 \ .
$}
\end{tabular}\mbox{}\\[2ex]

\subsection{The effective scalar-gluon-gluon coupling}
\label{app:heidi-hgg}

Straightforward evaluation of the matrix element for on-shell gluon fusion of
a single scalar mode $\phi_i$ via a top quark loop gives (\emph{cf.} \cite{Gunion:1989we})
\begin{multline}
\parbox{32mm}{\includegraphics{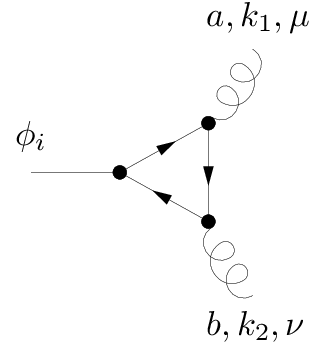}} +
\parbox{32mm}{\includegraphics{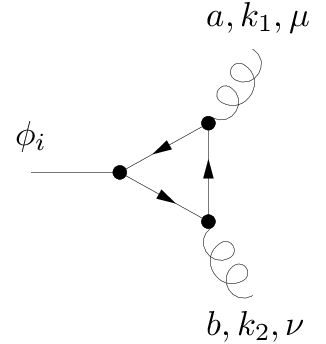}}
\quad = \\[2ex]
\frac{\alpha_s}{8\pi v}\xi^i_0\underbrace{
   \tau_i\left(1 + \left(1 - \tau_i\right)g(\tau_i)\right)}_{\rho(\tau_i)}
\;\left(k_1^\nu k_2^\mu - g^{\mu\nu}\left(k_1k_2\right)\right)
\;\delta_{ab}
\label{equ:gfusion-amp}
\end{multline}
where the function $g(\tau)$ is defined piecewise as
\begin{equation*}
g(\tau) = \begin{cases}
\arcsin^2\sqrt{\frac{1}{\tau}}
& \text{for }\tau \ge 1
\\
-\frac{1}{4}\left(\log\left(\frac{1+\sqrt{1-\tau}}{1-\sqrt{1-\tau}}\right)
   -i\pi\right)^2
& \text{for }\tau < 1
\end{cases}
\end{equation*}
and $\tau_i$ is defined as
\begin{equation*}
\tau_i = 4\frac{m_t^2}{\lambda_i^2}
\end{equation*}
with the top quark mass $m_t$ and the scalar field mass $\lambda_i$. The amplitude
Eq.~\eqref{equ:gfusion-amp} is reproduced by the trilinear part of the
gauge-invariant effective operator
\begin{equation}
\mathcal{O}_{gg\phi} = \sum_{i=0}^\infty\frac{\alpha_s}{16\pi v}\xi^i_0\rho(\tau_i)\;
\phi_i\tr G_{\mu\nu}G^{\mu\nu}
\label{equ:gfusion-op}
\end{equation}
with the gluon field strength tensor $G^{\mu\nu}$.

The scalar $\phi_i$ can be taken off the mass shell simply by replacing
$\lambda_i^2$ by its momentum $p^2$, and the lowest order of the expansion of
Eq.~\eqref{equ:gfusion-amp} in $p^2$ can then be obtained by taking the limit
\begin{equation*}
\lim_{\tau\rightarrow\infty}\rho(\tau) = \frac{2}{3}
\end{equation*}
Inserting this limit into Eq.~\eqref{equ:gfusion-op}, we obtain the lowest dimension
operator contributing to gluon fusion in the effective theory obtained by
consistently integrating out the top quark
\begin{equation}
\mathcal{O}_{gg\phi}^0 = \sum_{i=0}^\infty\frac{\alpha_s}{24\pi v}\xi^i_0\;
\phi_i\tr G_{\mu\nu}G^{\mu\nu}
\label{equ:gfusion-op-largemt}
\end{equation}

Both effective operators are available in
our {\sc FeynRules} HEIDI implementation via the function \texttt{LHEIDIgg}.
Calling this function as \texttt{LHEIDIgg["heavytop"]} generates
Eq.~\eqref{equ:gfusion-op-largemt}, while Eq.~\eqref{equ:gfusion-op} can be obtained by
omitting the argument. For the simulation results presented in this work,
Eq.~\eqref{equ:gfusion-op-largemt} has been used.


\clearpage

\end{appendix}

\end{document}